\documentclass[a4paper,11pt]{article}
\usepackage{jcappub} 
\usepackage{lineno}
\newcommand{\red}{\textcolor{black}}

\arxivnumber{2210.04000} 
\title{\boldmath Prospects for constraining interacting dark energy cosmology with gravitational-wave bright sirens detected by future FAST/SKA-era pulsar timing arrays}

\author[a]{Bo Wang,}
\emailAdd{bowang@nxu.edu.cn}
\affiliation[a]{School of Physics, Ningxia University, Yinchuan 750021, China}

\author[b,1]{Dong-Ze He,\note{Corresponding author.}}
\emailAdd{hedz@cqupt.edu.cn}
\affiliation[b]{College of Sciences, Chongqing University of Posts and Telecommunications, Chongqing 400065, China}

\author[c]{Ling-Feng Wang,}
\emailAdd{wanglf@hainanu.edu.cn}
\affiliation[c]{College of Physics and Optoelectronic Engineering, Hainan University, Haikou, 570228, China}

\author[d]{Hai-Li Li}
\affiliation[d]{Basic Teaching Department, Shenyang Institute of Engineering, Shenyang 110136, China}

\author[b]{and Yi Zhang}

\abstract{
We explore the constraints on cosmological parameters in interacting dark energy (IDE) models described by energy transfer rates $Q = \beta H \rho_{\rm de}$ and $Q = \beta H \rho_{\rm c}$, using simulated gravitational-wave (GW) bright siren data from pulsar timing arrays (PTAs) and the Planck 2018 cosmic microwave background (CMB) data. In particular, we simulate a future PTA observation in the FAST/SKA era with 20 millisecond pulsars (MSPs), each having 20\,ns white noise over a 10-year observation span, and demonstrate that this mock dataset significantly improves the constraint precision of key cosmological parameters such as the Hubble constant $H_0$, matter density $\Omega_m$, and the coupling parameter $\beta$. For the IDE model $Q = \beta H \rho_{\rm de}$, PTA data alone provides tighter constraints on these parameters than the CMB data alone, primarily due to the high sensitivity of GW standard sirens in probing the late universe. Combining PTA and CMB data further enhances the constraints by 43.6\% for $H_0$, 43.2\% for $\Omega_m$, and 44.7\% for $\beta$, relative to using CMB data alone. In contrast, for $Q = \beta H \rho_{\rm c}$, the CMB data alone constrains $\beta$ more tightly than the PTA data, due to the stronger impact of this interaction in the early universe. Nevertheless, the PTA+\,CMB combination still yields improvements of 13.3\% for $H_0$, 22.7\% for $\Omega_m$, and 18.2\% for $\beta$. Increasing the number of MSPs in the PTA further tightens all parameter constraints in both IDE models. Our results highlight the great potential of future PTA observations for significantly improving cosmological parameter estimation in IDE models, offering critical insights into the nature of dark energy and its interaction with dark matter.
}

\begin{document}
\maketitle
\flushbottom

\section{Introduction}
The discovery of the accelerating expansion of the universe through observations of Type Ia supernovae \cite{SupernovaCosmologyProject:1998vns, SupernovaSearchTeam:1998fmf}, later confirmed by cosmic microwave background (CMB) and large-scale structure observations \cite{WMAP:2003elm, WMAP:2003ivt, SDSS:2003eyi, SDSS:2004wzw}, led to the proposal of dark energy as a mechanism with negative pressure to explain this phenomenon \cite{Sahni:2006pa, Bamba:2012cp, Weinberg:1988cp, Peebles:2002gy, Copeland:2006wr, Frieman:2008sn, Sahni:2008zz, Li:2011sd, Kamionkowski:2007wv}. However, the fundamental nature of DE remains an open question. Various models have been proposed to describe dark energy, with the simplest being the cosmological constant ($\Lambda$) introduced by Einstein in 1917. The $\Lambda$ cold dark matter model, which includes $\Lambda$ as dark energy, aligns well with most cosmological observations, and recent datasets have constrained the relevant cosmological parameters to high precision \cite{Planck:2018vyg}.

Accurate measurements of cosmological parameters are crucial for understanding the universe's expansion and the nature of both dark energy and dark matter. Electromagnetic observations such as the CMB \cite{Planck:2018vyg}, supernovae type Ia \cite{Pan-STARRS1:2017jku}, and baryon acoustic oscillations \cite{BOSS:2016wmc, Beutler:2011hx, Ross:2014qpa} currently dominate these efforts. Gravitational waves (GWs), however, offer a unique avenue by directly determining luminosity distances from the waveform, positioning GW events as "standard sirens" in cosmology. When these GW events have electromagnetic counterparts, termed bright sirens, the corresponding redshifts can be measured. Binary neutron star mergers, typically associated with kilonovae and short gamma-ray bursts, serve as excellent GW sources with electromagnetic counterparts \cite{Dalal:2006qt, Nissanke:2009kt, Eichler:1989ve, Zhu:2021ram}, as demonstrated by the GW170817 event \cite{LIGOScientific:2017vwq, LIGOScientific:2017ync, LIGOScientific:2017zic}, which provided an initial measurement of the Hubble constant ($H_0$) with a 14\% precision \cite{LIGOScientific:2017adf}. With continued LIGO-Virgo observations, $H_0$ precision is expected to reach 2\% within five years \cite{Chen:2017rfc}. While stellar-mass binary black hole mergers generally lack electromagnetic counterparts (making them dark sirens), they still serve as valuable cosmological probes. Analysis of 47 sources from the LIGO–Virgo–KAGRA catalog achieved a $H_0$ measurement with 17\% precision \cite{LIGOScientific:2021aug}. Massive black hole binary mergers may also offer electromagnetic counterparts \cite{Palenzuela:2010nf, OShaughnessy:2011nwl, Moesta:2011bn, Kaplan:2011mz, Shi:2011us, Blandford:1977ds, Meier:2000wk, Dotti:2011um, Popovic:2011uy, DeRosa:2019myq, Bogdanovic:2021aav}, potentially improving $H_0$ estimates. Specifically, using massive black hole binary as bright and dark sirens with the Taiji-TianQin-LISA network could enhance $H_0$ precision to 0.9\% \cite{Jin:2023sfc}.

Supermassive binary black holes (SMBHBs), with masses $\geq 10^8 M_\odot$ residing in galactic centers, emit GWs in the nano-Hertz frequency range ($10^{-9}-10^{-6}$ Hz). These frequencies are particularly within the sensitivity range of pulsar timing arrays (PTAs). Currently, there are three principal PTA initiatives globally: the Parkes Pulsar Timing Array \cite{Hobbs:2013aka}, the European Pulsar Timing Array \cite{Kramer:2013kea}, and the North American Nanohertz Observatory for Gravitational Waves \cite{McLaughlin:2013ira}. These projects collaborate under the umbrella of the International Pulsar Timing Array \cite{Hobbs:2009yy}, aiming to boost detection sensitivities. While the primary focus has been on identifying the stochastic gravitational wave background (SGWB) \cite{Reardon:2023gzh,EPTA:2023fyk,NANOGrav:2023gor,Bian:2022tju,Bian:2023dnv}, research also extends to the detection of individual SMBHBs \cite{Rajagopal:1994zj,Jaffe:2002rt,Wyithe:2002ep,Sesana:2012ak,Ravi:2014aha,Perera:2018pts,Lee:2011et,Wang:2016tfy}. These SMBHBs, when utilized as standard sirens, offer a novel avenue to constrain cosmological parameters.

In Ref.\,\cite{Yan:2019sbx}, the authors performed a preliminary investigation on constraining the equation-of-state parameter of dark energy (with only equation-of-state parameter $w$ set free, the other cosmological parameters all fixed) using the SMBHBs expected to be detected by SKA-era PTAs, and found that the parameter can be constrained to an uncertainty of $\Delta w\sim 0.02-0.1$. Subsequently, in Ref.\,\cite{Wang:2022oou} the authors analyze the ability of SKA-era PTAs to detect existing SMBHB candidates in light of the simulation of timing residuals of pulsar signals, and use the mock data to constrain the cosmological parameters. They found that only 100 millisecond stable pulsars (MSPs) are needed to achieve precision cosmology if the root-mean-square (rms) of timing residuals could be reduced to 20 ns, and the SMBHB bright sirens could effectively break the cosmological parameter degeneracies inherent in the CMB, improving the constraint precision of the $w$ to $3.5\%$ level, which is comparable with the result of $Planck$ 2018.

However, the impact of SMBHB bright sirens on interacting dark energy (IDE) models, which allow for possible interactions between dark energy and dark matter, remains unexplored. In this work, we aim to fill this gap by analyzing the potential of PTAs to detect SMBHB GWs and by using mock data to constrain cosmological parameters within IDE models. This analysis will provide a more comprehensive understanding of how bright sirens from PTAs can improve the precision of cosmological parameter estimation.

In the non-interaction models, the energy density of each fluid component is conserved separately, $\dot{\rho}_i+3H(1 + \omega_i)\rho_i = 0$, where $H$ is the Hubble parameter, $w$ is the equation-of-state parameter, and the subscript $i=({\rm r,b,c,de})$ represents radiation, baryons, cold dark matter and dark energy, respectively. In contrast, IDE models modify the continuity equations for dark energy and cold dark matter to account for energy transfer:
\begin{align}
    &\dot{\rho}_{\rm de}+H(1+\omega)\rho_{\rm de}=-Q,\label{Inc1}\\
    &\dot{\rho}_c+H\rho_c=Q,\label{Inc2}
\end{align}
where $Q$ is the energy transfer rate between cold dark energy and dark matter. A positive $Q$ indicates cold dark matter decays into dark energy, a negative $Q$ indicates dark energy decays into cold dark matter, and $Q = 0$ implies no interaction. Here, we assume $w = -1$ for dark energy. Many forms of $Q$ have been proposed in the literature \cite{Amendola:1999er, Zhang:2005rj, Zhang:2005rg, Zhang:2004gc, Barrow:2006hia, He:2008tn, Valiviita:2009nu, Boehmer:2008av, Xia:2009zzb, Clemson:2011an, Li:2014eha, Li:2014cee, Zhang:2017ize, Cui:2015ueu, Valiviita:2015dfa, Guo:2017hea, Guo:2018ans, Yang:2018euj, Li:2019ajo, DiValentino:2019ffd, Xiao:2018jyl, Liu:2019ygl, Zhang:2021yof,Costa:2016tpb,Li:2024qso}. In this paper, we adopt a general phenomenological form of $Q=\beta H\rho_{\rm de}$ referred to as the DE-coupled model, and $ Q=\beta H\rho_{\rm c}$ referred to as the DM-coupled model, where $\beta$ is free dimensionless coupling parameters.

The structure of this paper is as follows: in Sec. \ref{meth}, we describe the methods for using PTAs to obtain SMBHB information and constrain cosmological parameters in IDE models. Sec. \ref{res} presents the results and analysis. Finally, we conclude in Sec. \ref{con}.

\section{METHODOLOGY}\label{meth}
\subsection{PTA and SMBHB}
GWs from SMBHBs induce timing residuals in MSP observations, which can be extracted by subtracting model-predicted times of arrival (ToAs) from observed ToAs. The rms of these residuals serves as a measure to detect and constrain GW signals.

For a GW source from direction $\hat{\Omega}$, the induced timing residual at time $t$ on Earth is given by \cite{Zhu:2015tua}:
\begin{equation}
    s(t,\hat{\Omega})=F_+(\hat{\Omega})\Delta A_+(t)+F_\times(\hat{\Omega})\Delta A_\times(t),\label{1}
\end{equation}
where $F_+(\hat{\Omega})$ and $F_\times(\hat{\Omega})$ are antenna pattern functions and are defined by \cite{Lee:2011et,Wahlquist:1987rx}
\begin{align}
F_+(\hat{\Omega})= &\frac{1}{4(1-\cos\theta)}\Big\{(1+\sin^2\delta)\cos^2\delta_{\rm p}\cos[2(\alpha-\alpha_{\rm p})] \nonumber\\
&-\sin 2\delta\sin2\delta_{\rm p}\cos(\alpha-\alpha_{\rm p})+\cos^2\delta(2-3\cos^2\delta_{\rm p})\Big\},\nonumber\\
F_\times(\hat{\Omega})=&\frac{1}{2(1-\cos\theta)}\Big\{\cos\delta\sin2\delta_{\rm p}\sin(\alpha-\alpha_{\rm p})\nonumber\\
&-\sin\delta\cos^2\delta_{\rm p}\sin[2(\alpha-\alpha_{\rm p})]\Big\}.\label{2}
\end{align}
Here, $(\alpha,\delta)$ and $(\alpha_{\rm p},\delta_{\rm p})$ are the right ascension and declination of the GW source and pulsar, respectively. $\theta$ is the angle between the GW source and pulsar with respect to the observer
\begin{equation}
    \cos\theta=\cos\delta\cos\delta_{\rm p}\cos(\alpha-\alpha_{\rm p})+\sin\delta\sin\delta_{\rm p}.\label{3}
\end{equation}
$\Delta A_{+/\times}(t)=A_{+/\times}(t)-A_{+/\times}(t_{\rm p})$ represents the difference between the Earth term and the pulsar term \cite{Jenet:2003ew}, with $t_{\rm p}=t-d_{\rm p}(1-\cos\theta)/c$ being the time the GW passes the MSP, where $d_{\rm p}$ is the pulsar distance. For a circular binary SMBHB system, these terms take the form \cite{Babak:2011mr, Ellis:2012zv}:
\begin{align}
    A_+(t)=&\frac{h_0(t)}{2\pi f(t)}\Big\{(1+\cos^2\iota)\cos2\psi\sin[\phi(t)+\phi_0]\nonumber+2\cos\iota\sin2\psi\cos[\phi(t)+\phi_0]\Big\},\\
    A_\times(t)=&\frac{h_0(t)}{2\pi f(t)}\Big\{(1+\cos^2\iota)\sin2\psi\sin[\phi(t)+\phi_0]\nonumber-2\cos\iota\cos2\psi\cos[\phi(t)+\phi_0]\Big\}.
\end{align}
Here, $\iota$ is the inclination angle of the binary orbit, $\psi$ is the GW polarization angle, $\phi_0$ is the phase constant. The GW strain amplitude $h_0(t)$ is defined as
\begin{equation}
    h_0(t)=2\frac{(GM_z)^{5/3}}{c^4}\frac{[\pi f(t)]^{2/3}}{d_{\rm L}},
\end{equation}
where $d_{\rm L}$ is the luminosity distance to the source; $M_z=M_c(1+z)$ is the redshifted chirp mass; $M_c=m_1^{3/5}m_2^{3/5}(m_1+m_2)^{-1/5}$ represents the binary chirp mass with $m_1$ and $m_2$ the SMBBH component masses.

The frequency and the orbital phase evolve according to
\begin{align}
    f(t)&=\left[f_0^{-8/3}-\frac{256}{5}\pi^{8/3}\left(\frac{GM_z}{c^3}\right)^{5/3}t\right]^{-3/8},\\
    \phi(t)&=\frac{1}{16}\left(\frac{GM_z}{c^3}\right)^{-5/3}\left\{(\pi f_0)^{-5/3}-[\pi f(t)]^{-5/3}\right\},
\end{align}
where $f_0=2f_{\rm orb}$ is the GW frequency at the time of the first observation. Here, $f_{\rm orb}=(2\pi T_S)^{-1}$ is the orbital frequency, in which the $T_S$ represents the orbital period of the SMBHBs.

For the GW sources, we use the current available 154 SMBHB candidates obtained from various characteristic signature. Most (149) of these samples are obtained via periodic variations in their light curves \cite{Graham:2015tba,Graham:2015gma,Charisi:2016fqw}, and the others are Mrk 231 \cite{Yan:2015mya}, NGC 5548 \cite{Li:2016hcm}, OJ 287 \cite{Valtonen:2008tx}, SDSS J0159+0105 \cite{Zheng:2015dij}, and Ark 120 \cite{Li:2017eqf}. These electromagnetic signals provide redshift information for the candidates, and if the corresponding GW signal is also detected, they can act as electromagnetic counterparts, making the SMBHB candidates bright sirens. For this analysis, we assume that the inclination angle $\iota$ is randomly distributed between $[0, \pi]$, and the polarization angle $\psi$ and initial phase $\phi_0$ are randomly chosen from $[0, 2\pi]$. To estimate the luminosity distances of these candidates, \red{we use their redshift information along with the CMB data from Planck 2018, which includes both the angular power spectra TT, TE, EE +lowE \cite{Planck:2018vyg} and lensing potential power spectrum reconstruction data from the Planck Public Release 4 (PR4)\cite{Carron:2022eyg}, to constrain the DE-coupled and DM-coupled models as the fiducial cosmology.}

The signal-to-noise ratio (SNR) of the GW signal detected by a PTA is defined as:
\begin{equation} \rho^2 = \sum_{j=1}^{N_{\rm p}} \sum_{i=1}^{N} \left[\frac{s_j(t_i)}{\sigma_{t,j}}\right]^2, \label{SNR}
\end{equation}
where $N_{\rm p}$ is the number of MSPs, $\sigma_{t,j}$ is the rms timing noise for the $j$-th MSP, and $N$ is the number of data points. The Fisher matrix to estimate the GW source parameters ${p}$ is given by:
\begin{equation}
F_{ab} = \sum_{j=1}^{N_{\rm p}} \sum_{i=1}^{N} \frac{\partial s_j(t_i)}{\sigma_{t,j}\partial p_a} \frac{\partial s_j(t_i)}{\sigma_{t,j}\partial p_b} - \frac{\partial^2 \ln P(p_c)}{\partial p_a \partial p_b},
\end{equation}
where $P(p_c)$ is the prior for the parameter $p_c$. The parameter set ${p}$ includes eight GW source parameters: ${M_c, \alpha, \delta, \iota, \psi, \phi_0, f_0, d_{\rm L}}$. Since the disk direction is randomly distributed in $4\pi$ solid angle, we need consider the prior of inclination angle $P(\iota)\propto\sin\iota$.

The ability of PTAs to detect individual GW source depends on several factors, including the number of MSPs, timing noise, observation span, and cadence. Following \cite{Yan:2019sbx}, we assume a 10-year observation period with a biweekly cadence. The rms timing noise $\sigma_t$ consists of two components: intrinsic white noise from practical measurements and the SGWB, which affects the timing residuals from the entire SMBHB population. Intrinsic white noise is expected to be significantly reduced with advancements from facilities like the Five-hundred-meter Aperture Spherical Telescope (FAST) \cite{Nan:2011um}, MeerKAT \cite{Bailes:2018azh}, and the Square Kilometre Array (SKA) \cite{Lazio:2013mea, Hobbs:2014tqa}. According to \cite{Porayko:2018sfa}, jitter noise will dominate the timing noise for 10 Parkes Pulsar Timing Array pulsars in the SKA era, with levels ranging from $10$ to $50\ {\rm ns}$. In this work, we construct three PTA samples for the FAST/SKA era with 20, 50, and 100 MSPs, referred to as PTA20, PTA50, and PTA100, respectively. These samples are selected using the Australia Telescope National Facility pulsar catalog \cite{Manchester:2004bp}, assuming an intrinsic white noise level of $20\ {\rm ns}$ for all ToAs. MSPs within 3 kpc of Earth are prioritized, as nearby pulsars generally have higher flux, leading to more accurate timing and higher SNR. For SGWB, some SMBHB population merger rate models can be used to describe the characteristic amplitude~\cite{Sesana:2008mz,Sesana:2008xk,Sesana:2016yky}. While in this work we consider a power law characteristic amplitude:
\begin{equation}
h_c(f)=A_{\rm GW}(\frac{f}{{\rm yr}^{-1}})^{-2/3},
\end{equation}
where we use $A_{\rm GW}=2.5\times10^{-15}$ following Ref.\,\cite{Ferranti:2024jsh} consistent with current observations \cite{Reardon:2023gzh,EPTA:2023fyk,NANOGrav:2023gor}. The corresponding one-sided power spectral density is given by $S_{\rm GW}(f) = h_c^2 / (12 \pi^2 f^3)$, and the resulting timing delay at frequency $f$ can be expressed as $\sqrt{S_{\rm GW}/T}$.

\begin{figure*}[!htb]
  	\begin{center}
  	\includegraphics[width=0.4\columnwidth]{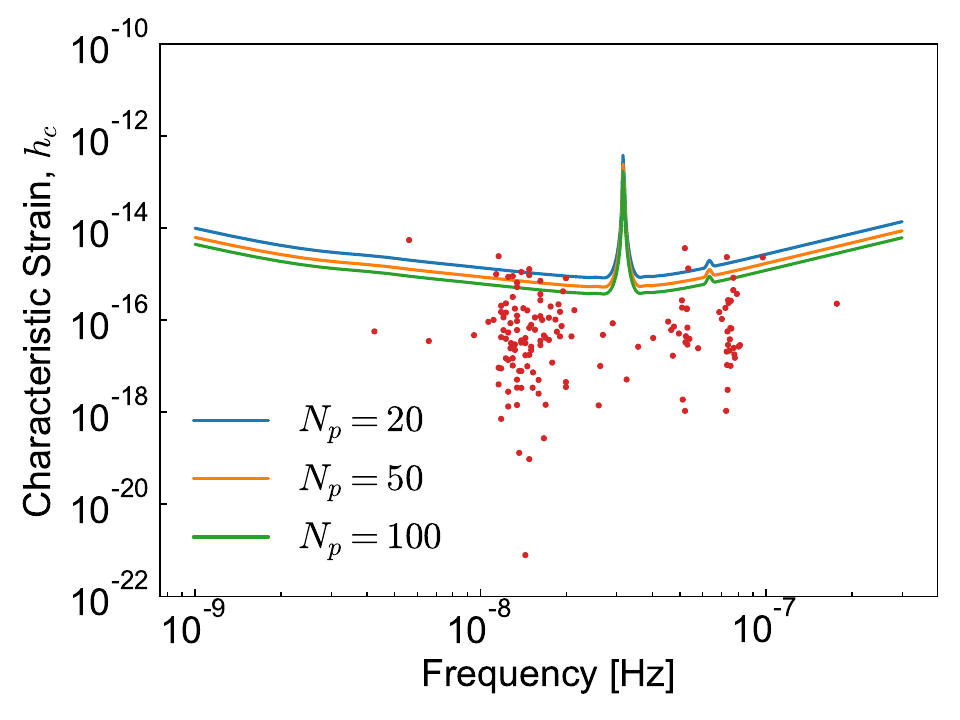}
   \includegraphics[width=0.48\columnwidth]{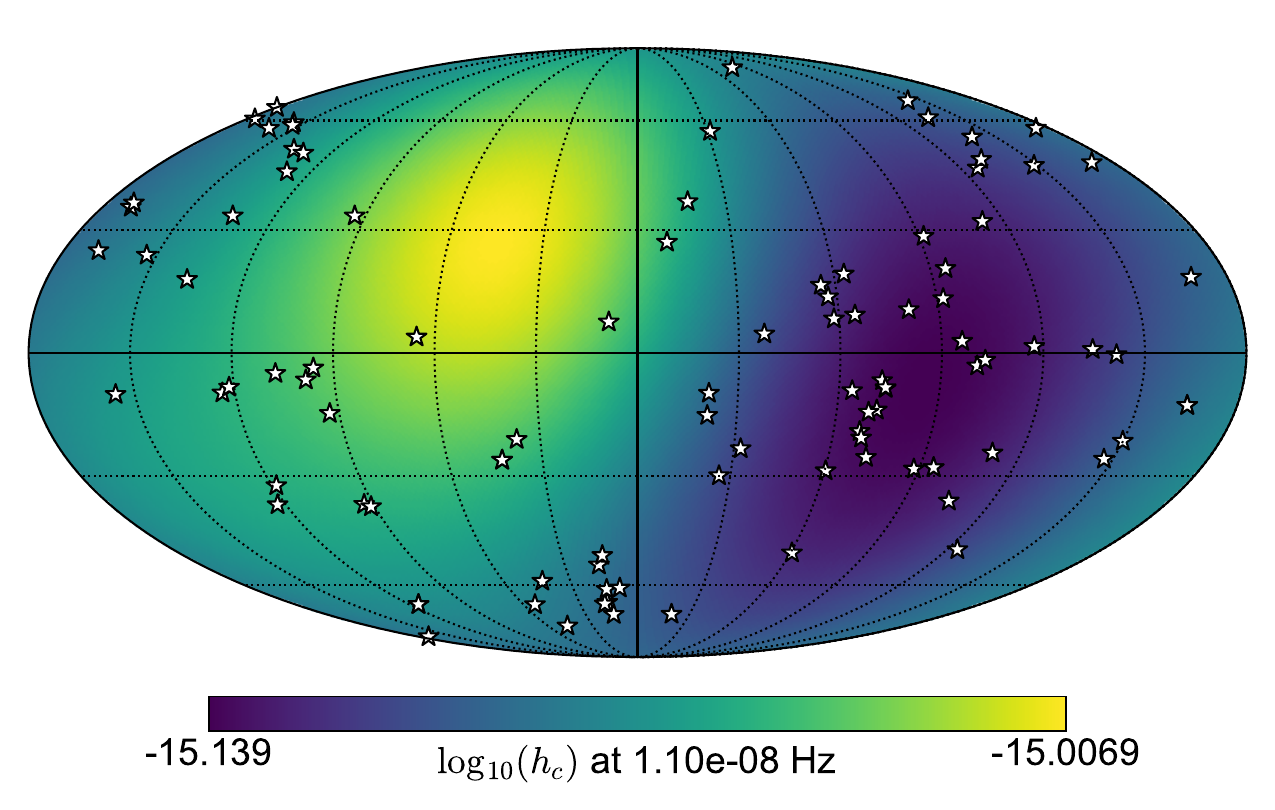}
  	\end{center}
  	\caption{Left: Sensitivity curves for FAST/SKA-era PTAs over a 10-year observation period, averaged over initial phase, inclination, and sky location. The curves represent cases with different numbers of MSPs, and the red dots indicate the GW strain amplitudes at $f = f_0$ for 154 SMBHB candidates. Right: All-sky sensitivity map at $f = 11$ nHz with 100 MSPs, where the white stars represent the pulsar locations.}
  	\label{SC}
  \end{figure*}

It is crucial to distinguish between individual GW signals from SMBHBs and the SGWB, as both contribute to PTA observations \cite{Ferranti:2024jsh}. We calculate the sensitivity curves using the ${\tt hasasia}$ package \cite{Thrane:2013oya, Moore:2014lga, Hazboun:2019vhv} based on the simulated PTA configurations and noise model. The results, averaged over inclination, polarization, and sky position, are shown in the left panel of Fig. \ref{SC}, with red dots representing the 154 SMBHB candidates. As the number of MSPs $N_p$ increases, the detection capability improves significantly. Identifying the GW source’s sky location is also important \cite{EPTA:2015gke, Mingarelli:2017fbe}. By calculating the sensitivity curves without averaging over sky position, we generate an all-sky GW strain sensitivity map at any given frequency (the right panel of Fig. \ref{SC} shows the map at 11 nHz based on 100 pulsars as an example), enabling a direct comparison between the SMBHB strain and the sensitivity at the same frequency and location to evaluate detectability.

\begin{figure}[!htp]
  	\begin{center}
  	\includegraphics[width=0.8\columnwidth]{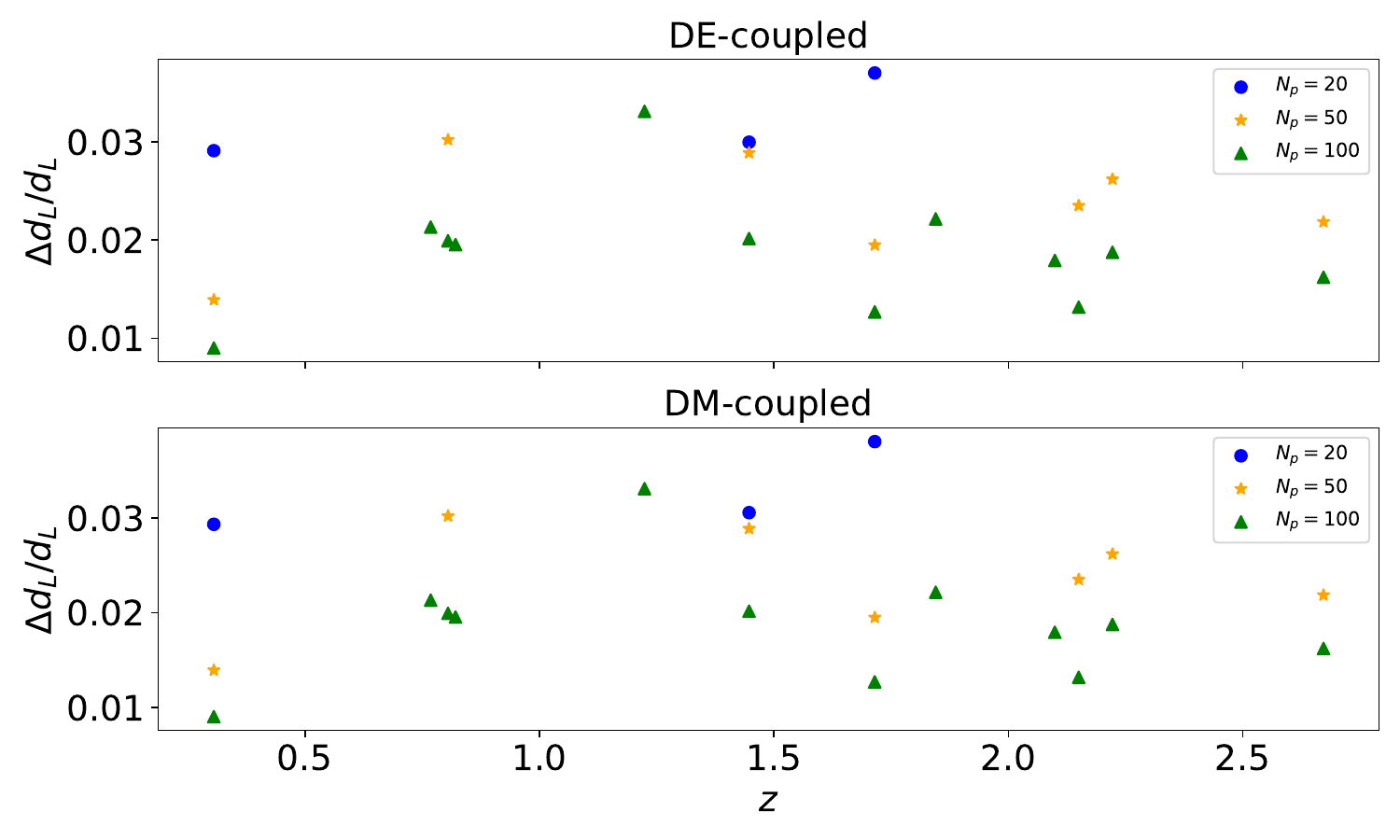}
  	\end{center}
  	\caption{Simulated precision of the luminosity distances ($\Delta d_{\rm L}/d_{\rm L}$) for SMBHB candidates with SNR $> 10$, based on the Fisher matrix analysis using 20, 50, and 100 MSPs. The luminosity distances $d_{\rm L}$ are calculated from the $Planck$ 2018 best-fit DE-coupled and DM-coupled models, with the absolute error $\Delta d_L$ being 1$\sigma$ confidence level.}
  	\label{DL}
  \end{figure}

\subsection{Cosmological Parameter Estimation}
We first calculate the SNR of different GW sources using Eq. (\ref{SNR}), and then apply the Fisher matrix method described in the previous subsection to obtain constraints on the luminosity distances of the SMBHBs. Only SMBHBs with an SNR greater than 10 are considered viable candidates for this analysis. For both the DE-coupled and DM-coupled models, we find that 15, 17, and 20 SMBHBs meet the SNR criterion for PTAs composed of 20, 50, and 100 MSPs, respectively. After comparing these candidates with the sensitivity curves, we retain only 3, 7, and 12 SMBHBs for further analysis. The simulated accuracies of the luminosity distances ($\Delta d_{\rm L}/d_{\rm L}$) for the retained SMBHBs are shown in Fig. \ref{DL}. As expected, for a given SMBHB candidate, increasing the number of MSPs results in tighter constraints on the luminosity distance $d_{\rm L}$.

\red{Next, we use the simulated data to constrain cosmological parameters through a Markov Chain Monte Carlo (MCMC) analysis. For a comprehensive analysis and comparison, we also incorporate the Planck 2018 dataset and the Planck PR4 lensing dataset which label as CMB}. To address the issue of perturbation divergence in IDE models, we adopt the extended parameterized post-Friedmann (ePPF) framework for handling cosmological perturbations \cite{Li:2014cee, Li:2014eha}. The MCMC calculations are performed using the ${\tt CosmoMC}$ package \cite{Lewis:2002ah}, while the IDE background dynamics and linear perturbation equations are implemented under a modified version of ${\tt CAMB}$ \cite{Lewis:1999bs}, with the ePPF code integrated to resolve perturbation divergence in the global cosmological fit.

\section{Result}\label{res}

   \begin{figure*}[!htb]
  	\begin{center}
  	\includegraphics[width=0.4\columnwidth]{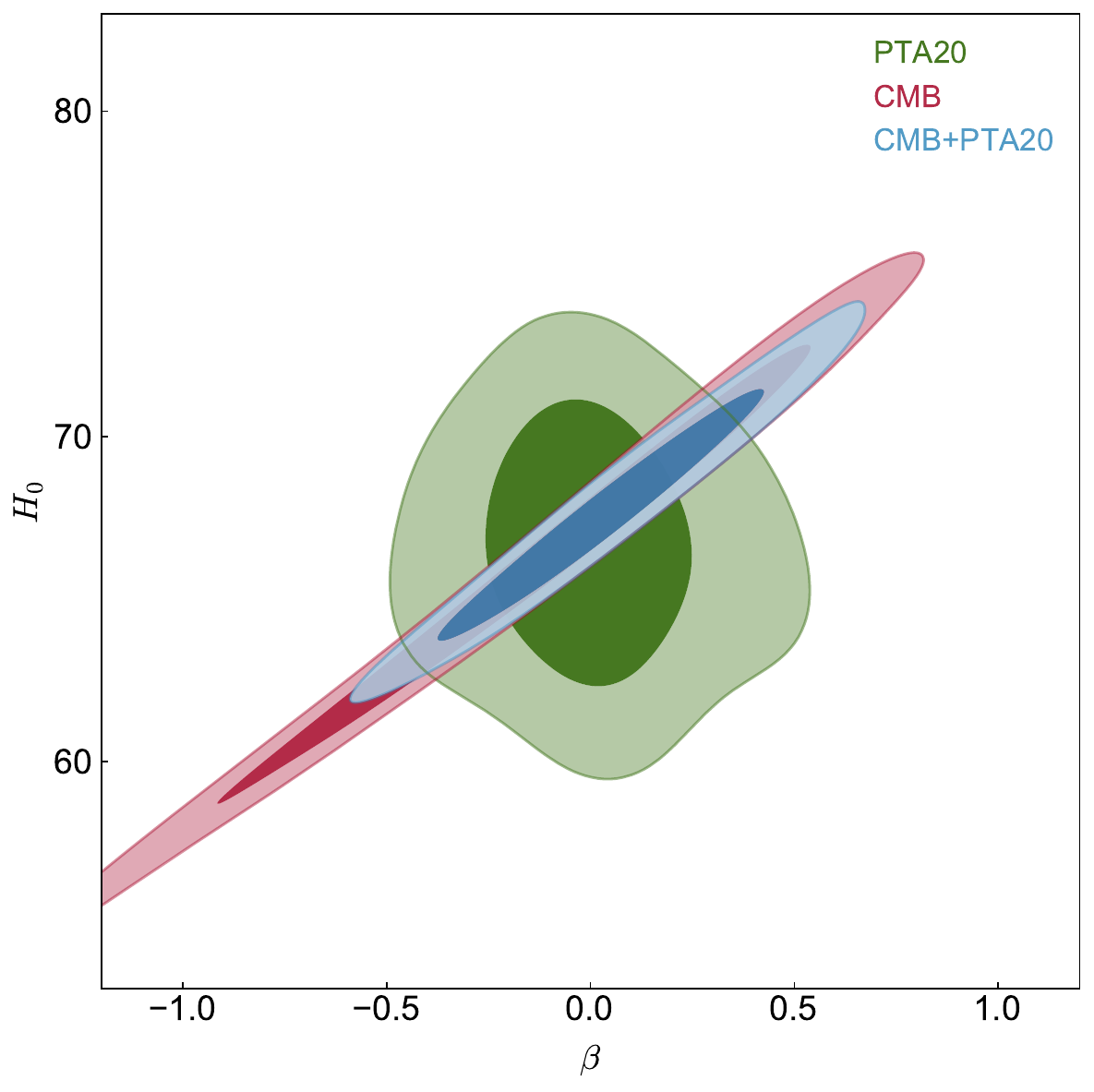}
   \includegraphics[width=0.4\columnwidth]{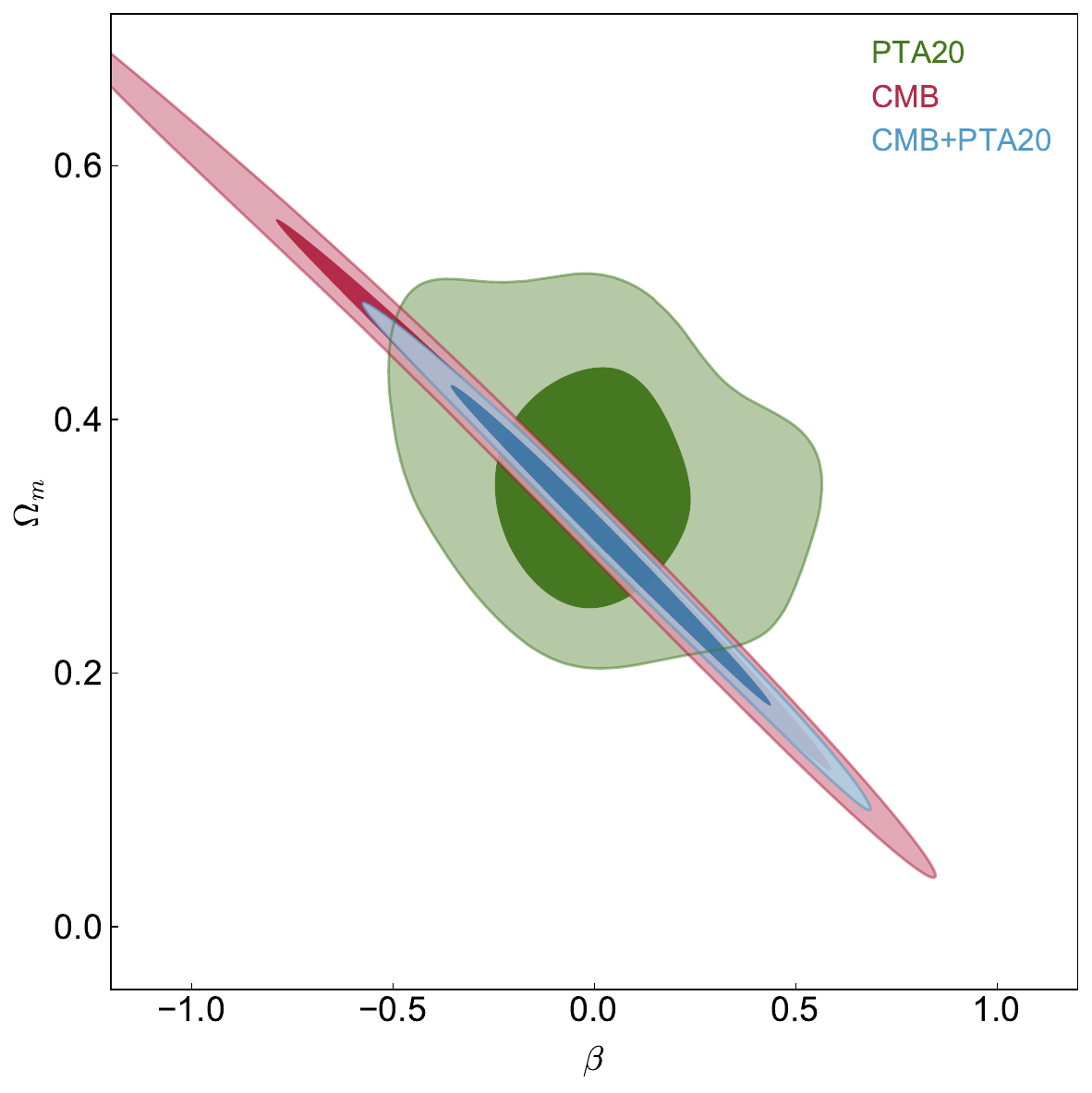}
  	\end{center}
  	\caption{Constraints on the cosmological parameters in the DE-coupled model from PTA20, CMB, and CMB+PTA20, respectively.}
  	\label{ILCDM1_1}
  \end{figure*}

   \begin{figure*}[!htb]
  	\begin{center}
  	\includegraphics[width=0.4\columnwidth]{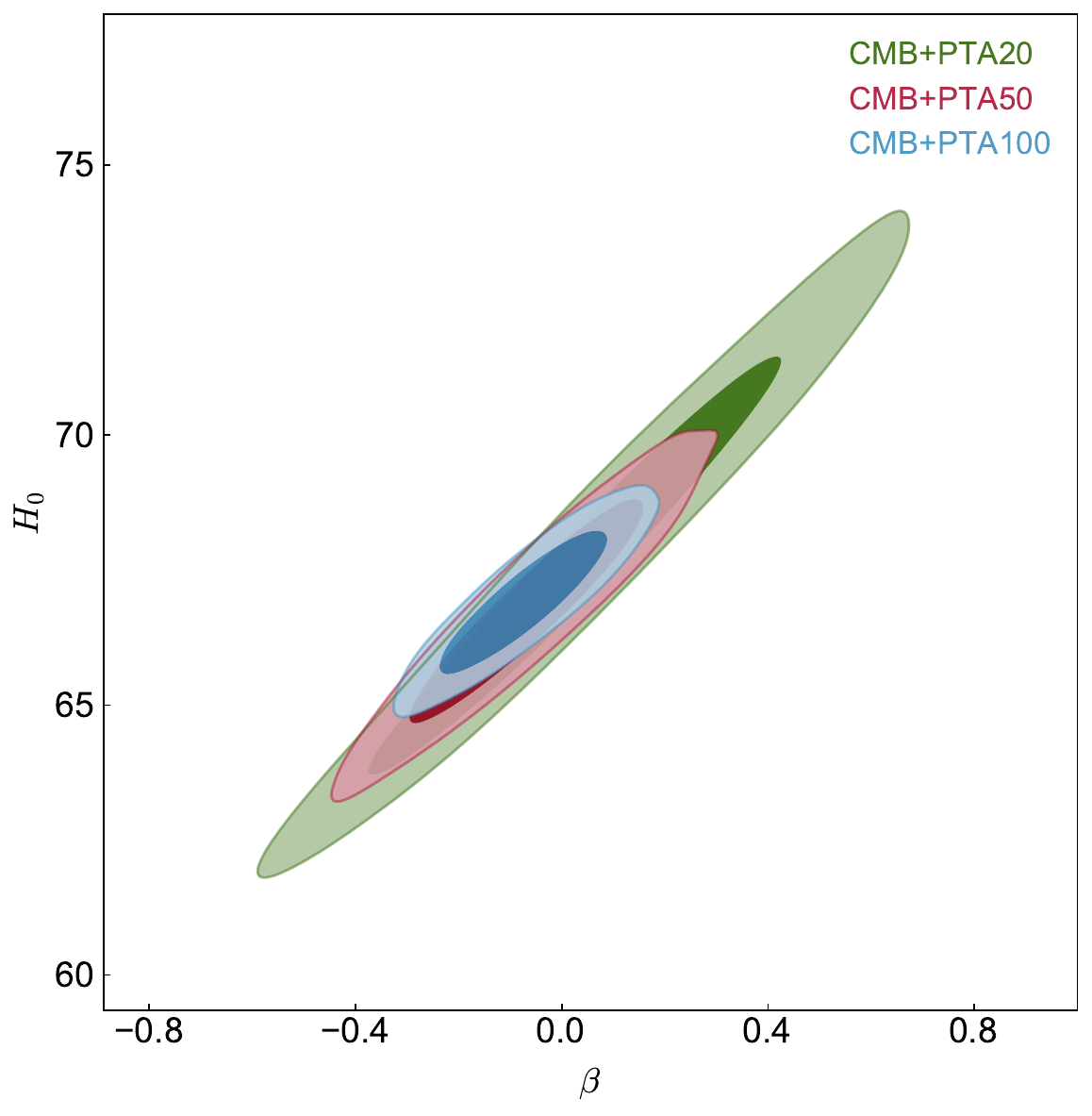}
   \includegraphics[width=0.4\columnwidth]{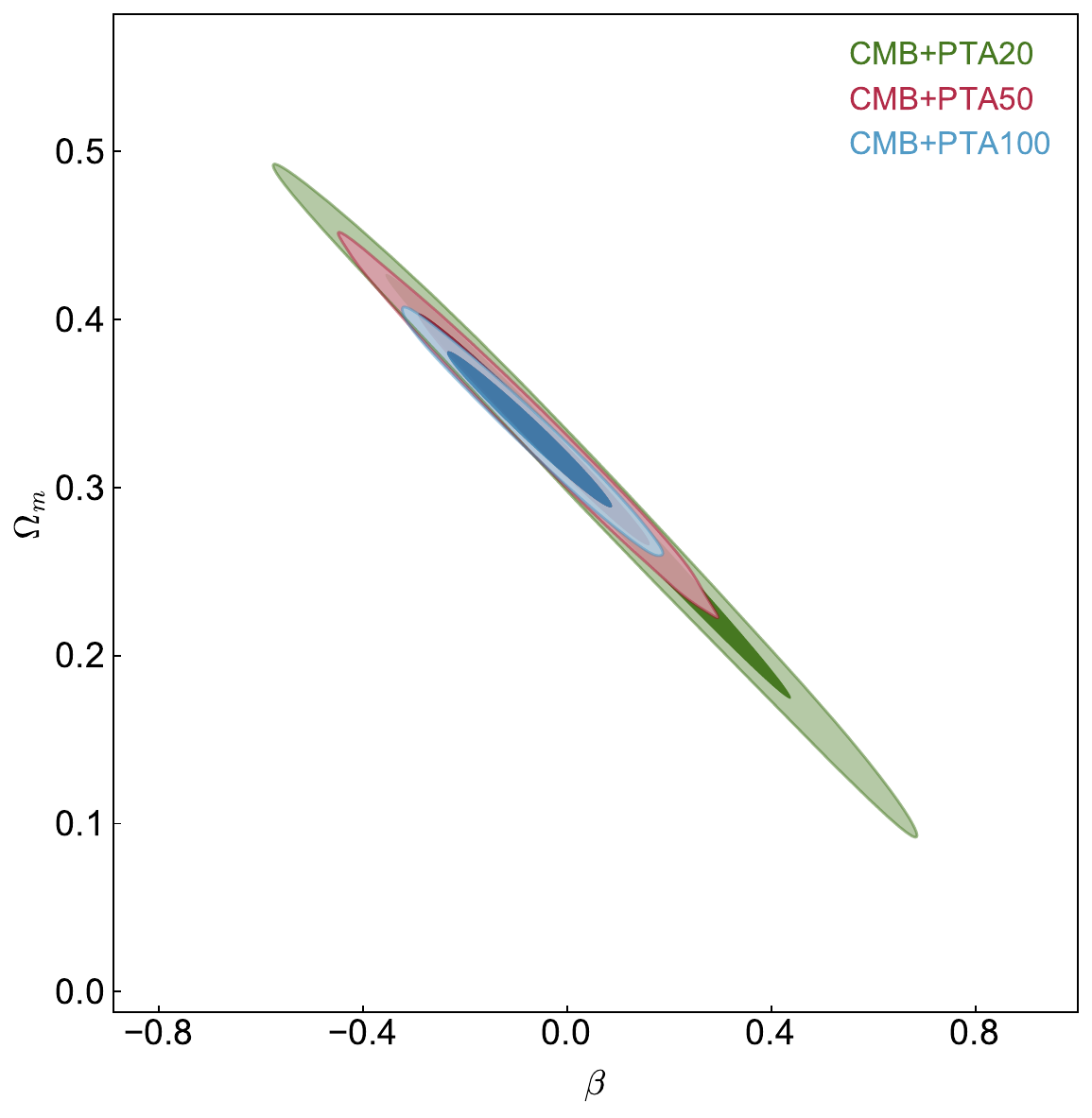}
  	\end{center}
  	\caption{Constraints on the cosmological parameters in the DE-coupled model from CMB+PTA20, CMB+PTA50, and CMB+PTA100, respectively.}
  	\label{ILCDM1_2}
  \end{figure*}

\begin{table}[!htp]
\centering
\begin{tabular}{l|c|c|c}
\hline
Data&$\sigma(\beta)$& $\sigma(H_0$)&$\sigma(\Omega_m)$ \\
\hline
CMB        & 0.47 & 4.54 & 0.146 \\
PTA20        & 0.15 & 2.40 & 0.055 \\
CMB+PTA20 & 0.26 & 2.56 & 0.083 \\
CMB+PTA50 & 0.15 & 1.40 & 0.046 \\
CMB+PTA100 & 0.11 & 0.87 & 0.031 \\
\hline
\end{tabular}
\caption{The $1\sigma$ absolute errors on the parameters in the DE-coupled model from different data combinations. Here, $H_0$ is in units of ${\rm km} \ {\rm s}^{-1} {\rm Mpc}^{-1}$.\label{ILCDM1}}
\end{table}

\begin{figure*}[!htb]
  	\begin{center}
  	\includegraphics[width=0.4\columnwidth]{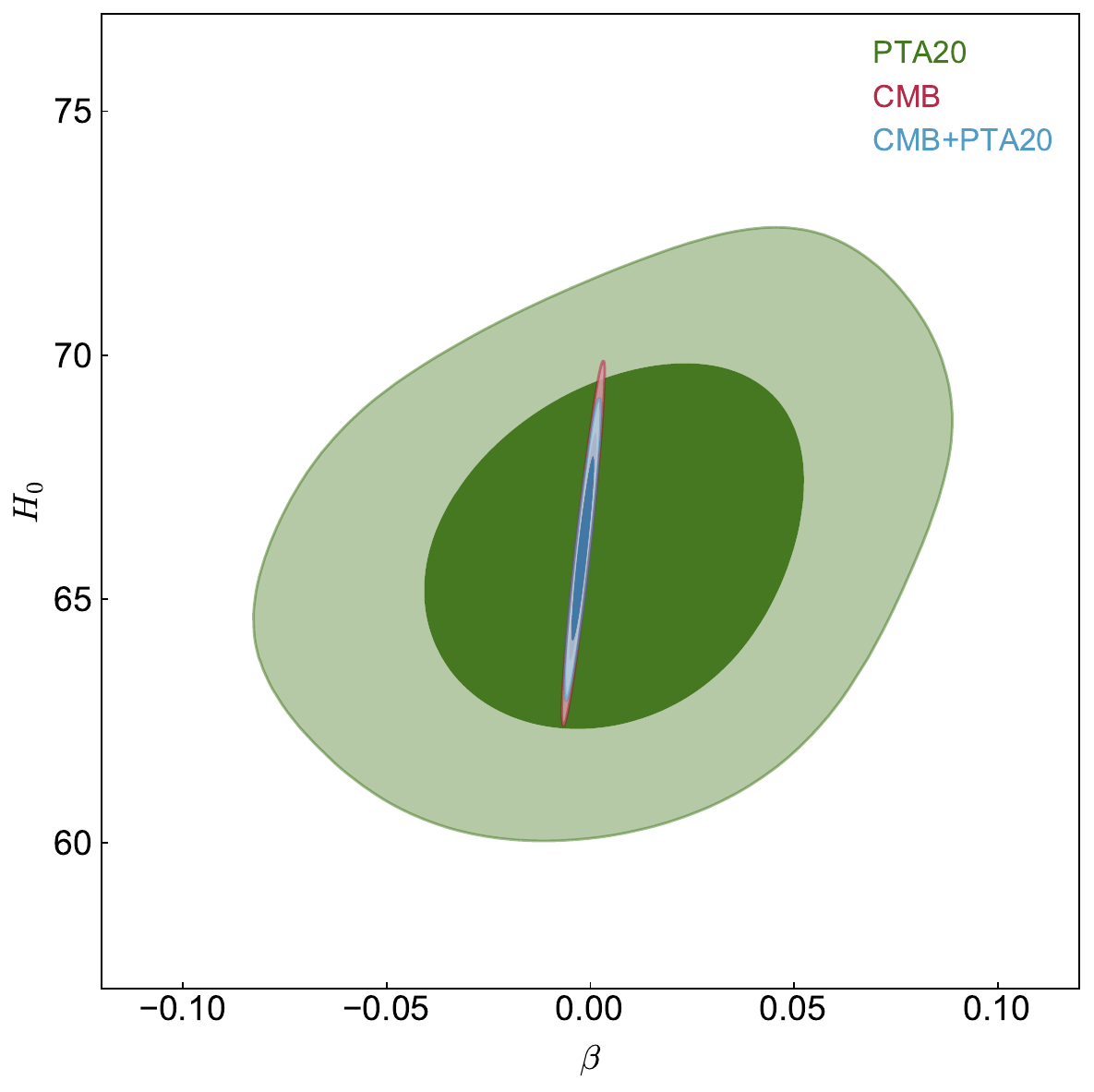}
   \includegraphics[width=0.4\columnwidth]{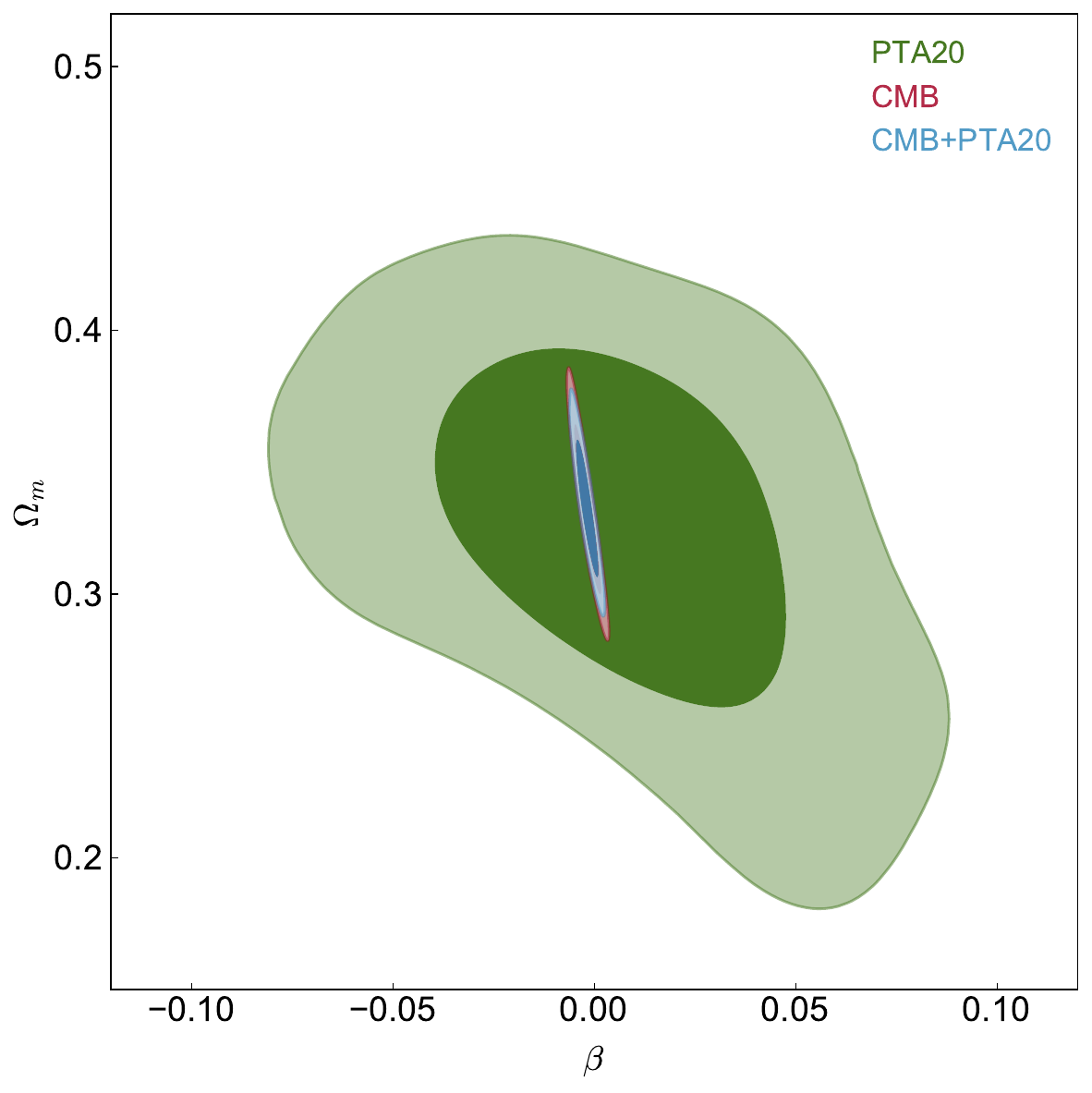}
  	\end{center}
  	\caption{Constraints on the cosmological parameters in the DM-coupled model from PTA20, CMB, and CMB+PTA20, respectively.}
  	\label{ILCDM2_1}
  \end{figure*}

\begin{figure*}[!htb]
  	\begin{center}
  	\includegraphics[width=0.4\columnwidth]{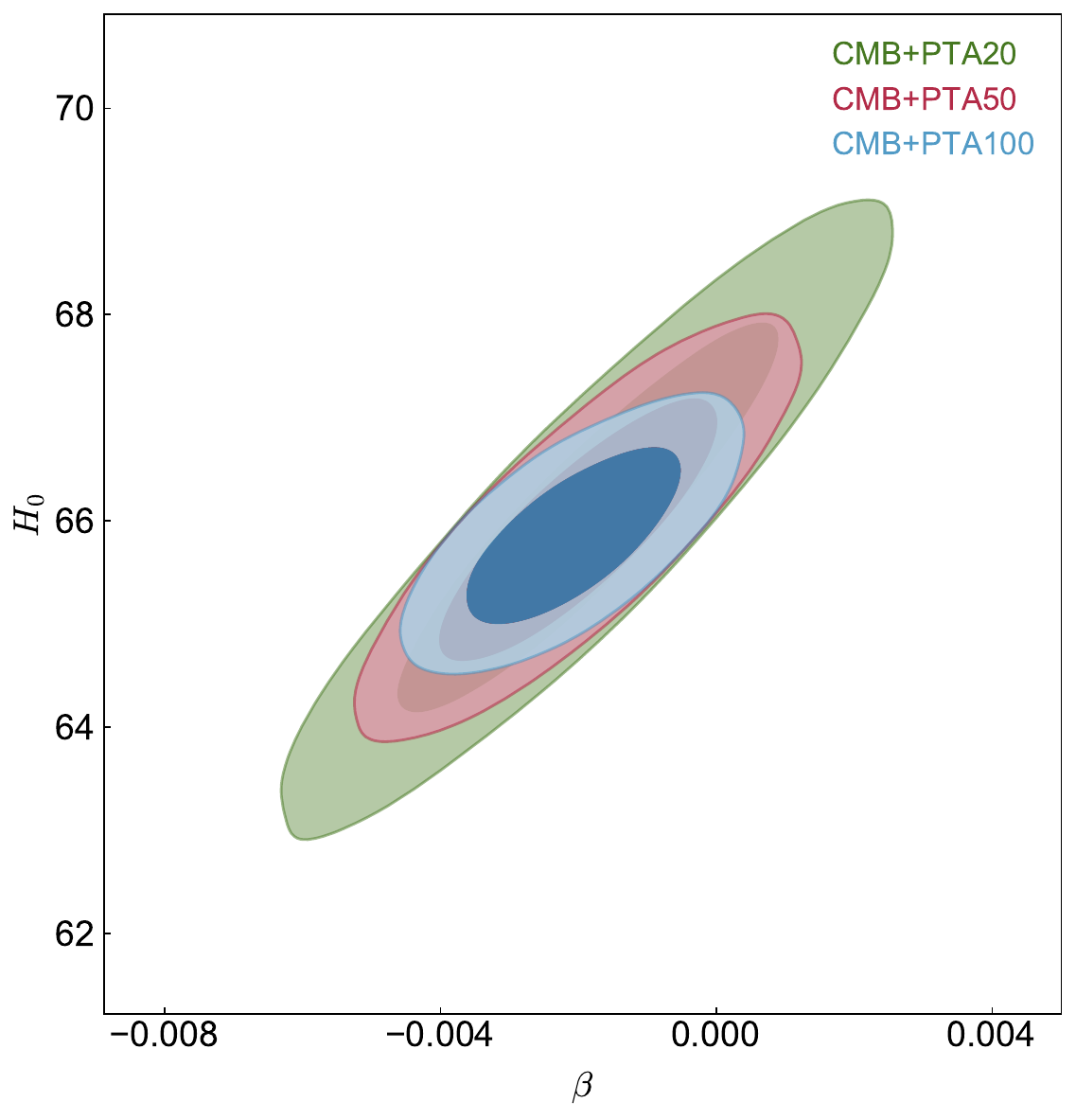}
   \includegraphics[width=0.4\columnwidth]{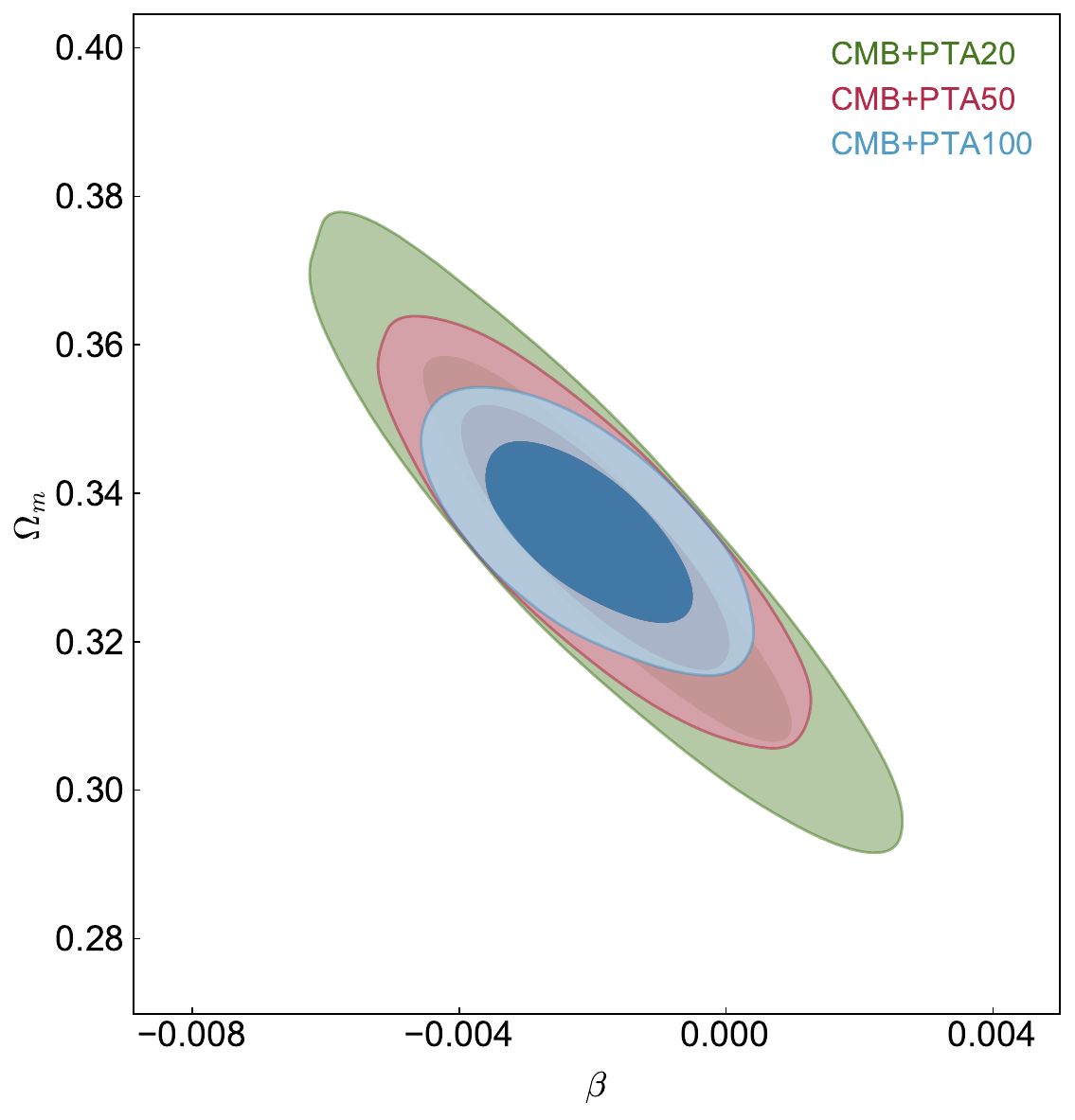}
  	\end{center}
  	\caption{Constraints on the cosmological parameters in the DM-coupled model from CMB+PTA20, CMB+PTA50, and CMB+PTA100, respectively.}
  	\label{ILCDM2_2}
  \end{figure*}

In this section, we shall report the constraint results on the models of DE-coupled and DM-coupled from the CMB data, simulated PTA data, and the data combinations of CMB and PTA. The constraint results are shown in Figs.\,\ref{ILCDM1_1}--\ref{ILCDM2_2}, and summarized in Tables\,\ref{ILCDM1}--\ref{ILCDM2}. In the following, for a cosmological parameter $\xi$, we use the $\sigma(\xi)$ to represent its $1\sigma$ absolute errors.

Fig.~\ref{ILCDM1_1} shows the $1\sigma$ and $2\sigma$ posterior distribution contours for the parameters in the DE-coupled model using PTA20, CMB, and CMB+PTA20 data combination. At first glance, it is evident that the CMB and simulated PTA data have different degeneracy orientations for the DE-coupled model. As a result, combining the CMB and PTA data provides tighter parameter constraints compared with using the CMB data alone. Using the PTA20 data alone, we achieve $\sigma(H_0) = 2.40\ {\rm km} \ {\rm s}^{-1} {\rm Mpc}^{-1}$ and $\sigma(\Omega_m) = 0.055$, which are significantly better than the constraint results from CMB data alone: $\sigma(H_0) = 4.54\ {\rm km} \ {\rm s}^{-1} {\rm Mpc}^{-1}$ and $\sigma(\Omega_m) = 0.146$. By combining the CMB and PTA20 data, we obtain $\sigma(H_0) = 2.56\ {\rm km} \ {\rm s}^{-1} {\rm Mpc}^{-1}$ and $\sigma(\Omega_m) = 0.083$. Compared with the CMB-only case, the CMB+PTA20 data combination could improve the constraint accuracy on $H_0$ and $\Omega_m$ by 43.6\% and 43.2\% respectively.

For the coupling parameter $\beta$ in the DE-coupled model, the CMB data alone provides a relatively weak constraint of $\sigma(\beta) = 0.47$. However, the PTA20 data alone gives a much tighter constraint, with $\sigma(\beta) = 0.15$. This is because, the CMB data is originated from the early universe, but DE mainly dominates the evolution of late-time universe. Thus, the GW data from PTA, as a late-universe probe,  is more sensitive to the effects of coupling between dark energy and dark matter in the IDE model with $Q = \beta H \rho_{\rm de}$. Consequently, the PTA data offer better sensitivity to $\beta$ compared to the CMB data, which primarily reflects early-universe conditions. The combination of CMB+PTA20 further improves the accuracy, yielding $\sigma(\beta) = 0.26$, which is a 44.7\% improvement over the CMB-only case.

Fig.~\ref{ILCDM1_2} shows the $1\sigma$ and $2\sigma$ measurement error contours for the DE-coupled model parameters using the CMB data combined with PTA data sets containing different numbers of MSPs. It is evident that as the number of MSPs  increases, the constraints become progressively tighter. For example, we obtain $\sigma(H_0) = 1.40\ {\rm km} \ {\rm s}^{-1} {\rm Mpc}^{-1}$ for CMB+PTA50 and $\sigma(H_0) = 0.87\ {\rm km} \ {\rm s}^{-1} {\rm Mpc}^{-1}$ for CMB+PTA100, improving the constraint accuracies by 45.3\% and 66.0\%, respectively, compared with the CMB+PTA20 case. For $\Omega_m$, we obtain $\sigma(\Omega_m) = 0.046$ for CMB+PTA50 and $\sigma(\Omega_m) = 0.031$ for CMB+PTA100, which are 44.6\% and 62.7\% improvements respectively, compared with the CMB+PTA20 case. For the coupling parameter $\beta$, the CMB+PTA50 and CMB+PTA100 combinations yield $\beta = 0.15$ and $\beta = 0.11$, improving the constraint accuracies by 42.3\% and 57.7\%, respectively, compared to the CMB+PTA20 case.

\begin{table}[!htp]
\centering
\begin{tabular}{l|c|c|c}
\hline
Data& $\sigma(\beta)$  & $\sigma(H_0)$ & $\sigma(\Omega_m)$ \\
\hline
CMB        & 0.0022 &  1.50 & 0.022 \\
PTA20      & 0.0265 &  2.20 & 0.039 \\
CMB+PTA20 & 0.0018 &  1.30 & 0.017 \\
CMB+PTA50 & 0.0013 &  0.84 & 0.012 \\
CMB+PTA100 & 0.0010 &  0.56 & 0.008 \\
\hline
\end{tabular}
\caption{The $1\sigma$ absolute errors on the parameters in the DM-coupled model from different data combinations. Here, $H_0$ is in units of ${\rm km} \ {\rm s}^{-1} {\rm Mpc}^{-1}$.\label{ILCDM2}}
\end{table}

The constraint results for the DM-coupled model with $ Q = \beta H \rho_{\rm c}$ are presented in Table~\ref{ILCDM2} and Figs.~\ref{ILCDM2_1}--\ref{ILCDM2_2}. These results differ significantly from the DE-coupled model. For the parameters $H_0$ and $\Omega_m$, using the PTA20 data alone, we obtain $\sigma(H_0) = 2.20\ {\rm km} \ {\rm s}^{-1} {\rm Mpc}^{-1}$ and $\sigma(\Omega_m) = 0.039$, which are less accurate than the results from the CMB data alone, $\sigma(H_0) = 1.50\ {\rm km} \ {\rm s}^{-1} {\rm Mpc}^{-1}$ and $\sigma(\Omega_m) = 0.022$. However, when combining the CMB and PTA20 data, the constraints could be improved, yielding $\sigma(H_0) = 1.30\ {\rm km} \ {\rm s}^{-1} {\rm Mpc}^{-1}$ and $\sigma(\Omega_m) = 0.017$. Compared to the CMB-alone case, the CMB+PTA20 data combination improves the constraints on $H_0$ and $\Omega_m$ by 13.3\% and 22.7\%, respectively.

For the coupling parameter $\beta$ in the DM-coupled model, the results are completely opposite to those in the DE-coupled model. The CMB data alone provides the tightest constraint on $\beta$, with $\sigma(\beta) = 0.0022$, which is significantly better than the result from PTA20 data alone, $\sigma(\beta) = 0.0265$. This is because, in the IDE model with $Q = \beta H \rho_{c}$, both the Hubble parameter $H$ and cold dark matter density $\rho_{\rm c}$ take much higher values in the early universe, allowing the energy transfer rate $Q$ to reach moderate values even for rather small value of $\beta$. Thus, the CMB data, as an early-universe probe, could provide a much tighter constraint on $\beta$ compared to PTA data, which is a late-universe probe. When combining the CMB and PTA20 data, we obtain $\sigma(\beta) = 0.0018$, an improvement of 18.2\% over the CMB-alone case.

The constraint results for the DM-coupled model from the data combination of CMB and PTA are shown in Fig.~\ref{ILCDM2_2}. As the number of MSPs in PTA data increases, the constraints on $H_0$, $\Omega_m$, and $\beta$ are steadily improved. The absolute errors for $H_0$ are $\sigma(H_0) = 0.84$ and $\sigma(H_0) =0.56$ for CMB+PTA50 and CMB+PTA100, respectively, showing the improvements of 35.4\% and 56.9\% compared to the CMB+PTA20 case. Similarly, for $\Omega_m$, we find $\sigma(\Omega_m) = 0.012$ for CMB+PTA50 and $\sigma(\Omega_m) =0.008$ for CMB+PTA100, improved by 29.4\% and 52.9\%, respectively, compared with the CMB+PTA20 case. For the coupling parameter $\beta$, the constraint results using CMB+PTA50 and CMB+PTA100 are $\sigma(\beta) = 0.0013$ and $\sigma(\beta) =0.0010$ respectively, showing the improvements of 27.8\% and 44.4\% compared to the CMB+PTA20 case.

\section{Conclusion}\label{con}

In this work, we investigated the constraints on cosmological parameters in the IDE models, namely DE-coupled and DM-coupled, using simulated PTA data in the FAST/SKA era alongside the CMB data from Planck 2018. It is shown that the inclusion of future PTA data could significantly enhance the constraint precision of cosmological parameters.

Using the upcoming FAST/SKA experiment as a reference, we simulated three PTA data sets, consisting of 20, 50, and 100 MSPs with 20 ns white noise. Each PTA data sets include the consideration of a 10-year observation span with bi-weekly ToA measurements. The SMBHB candidates with SNR $>$ 10 are selected in our simulation, and we found that increasing the number of pulsars in a PTA leads to a higher number of detectable SMBHB candidates and significantly improves the accuracy of their luminosity distances.

For the DE-coupled model, the PTA data could provide tighter constraints on the parameters $H_0$, $\Omega_m$, and the coupling parameter $\beta$ compared with the CMB data, due to the higher sensitivity of GW observations in probing the late universe in which the effects of DE are more pronounced. With the combination of the PTA and CMB, the constraints on these parameters can be further improved. By combining the CMB data with the PTA20 data, the constraints on $H_0$ and $\Omega_m$ are improved by 43.6\% and 43.1\%, respectively, and the constraint on $\beta$ is improved by 44.4\%. With the increase of the number of MSPs in the PTA, much tighter constraints can be obtained. When using the CMB+PTA100 data, the constraint errors are improved by 66.0\% for $H_0$, 62.6\% for $\Omega_m$, and 57.8\% for $\beta$, compared with the case using CMB+PTA20.

For the DM-coupled model, the results were notably different. The CMB data could provide a tighter constraint on the coupling parameter $\beta$ compared with the PTA data, due to the significant influence of the coupling on the early universe. However, combining the CMB data with PTA data still led to the improvement of parameter constraints. The data combination of CMB+PTA20 could improve the constraints on $H_0$, $\Omega_m$, and $\beta$ by 13.3\%, 21.3\%, and 18.2\%, respectively. With the increase of the MSPs number in PTA, the parameter constraints can be further tightened. When using the CMB+PTA100 data, the constraint errors are improved by 56.9\% for $H_0$, 52.9\% for $\Omega_m$, and 44.4\% for $\beta$, compared with the case using CMB+PTA20.

\red{Finally, we emphasize that PTAs serve as an independent late-time cosmological probe, complementing other late-time observables such as large-scale structure and supernovae surveys. Although these future surveys will undoubtedly provide stronger constraints, PTAs bring distinct systematics and observational techniques, thereby helping to break degeneracies that may remain in other probes. In this work, we have illustrated how combining simulated PTA data with real CMB data can substantially tighten parameter constraints in IDE models. Once actual PTA observations become available, it will be possible to cross-check for any tensions or systematic effects among various datasets, further advancing our understanding of the dark sector.}

Overall, our results demonstrate that the future GW observations with PTA have great potential to significantly improve the precision of cosmological parameter estimation in the IDE cosmology. This highlights the important role that GW observations with PTA will play in future cosmological studies, particularly in constraining the nature of dark energy and the interactions between dark energy and dark matter.

\acknowledgments

We thank Yue Shao for helpful discussions. Dong-Ze He is supported by the Talent Introduction Program of Chongqing University of Posts and Telecommunications (grant No. E012A2021209), the Youth Science and technology research project of Chongqing Education Committee (Grant No. KJQN202300609). Ling-Feng Wang is supported by the National Natural Science Foundation of China (Grant No.12305058), and the Natural Science Foundation of Hainan Province of China (Grant No.424QN215). Hai-Li Li is supported by the National Natural Science Foundation of China (Grant No.12305068), and the Natural Science Foundation of Liaoning Province of China (Grant No.2023-BSBA-229). Yi Zhang is supported by the National Natural Science Foundation of China (Grant No.12275037), and the CQ CSTC (Grant Nos. cstc2020jcyj-msxmX0810 and cstc2020jcyj-msxmX0555).


\bibliographystyle{JHEP}
\bibliography{refs}




\end{document}